# Observations of Ultrafast Kelvin Wave Breaking in the Mars Thermosphere.

**E.M.B. Thiemann**[1], N. Entin[1,2], S. Bougher[3], E. Yiğit[4], J. Bell[5], D. Pawlowski[6], F. Eparvier[1]

(1) Laboratory for Atmospheric and Space Physics, University of Colorado at Boulder (thiemann@lasp.colorado.edu); (2) Peak to Peak Charter School, Lafayette CO; (3) Department of Climate and Space Sciences and Engineering, University of Michigan at Ann Arbor; (4) Department of Physics, George Mason University; (5) NASA Goddard Space Flight Center, Greenbelt MD; (6) Eastern Michigan University at Ypsilanti

**Introduction:** Ultrafast Kelvin Waves (UFKWs) have been recently discovered at Mars using measurements by the Mars Atmosphere and Evolution (MAVEN) Accelerometer (ACC) instrument and the Mars Reconnaissance Orbiter (MRO) Mars Climate Sound (MCS) instrument [1]. UFKWs are eastward propagating, have a 2-3 sol period with sufficiently long vertical wavelengths to propagate into the thermosphere, and are predicted by classical wave theory to be equatorially trapped. These prior measurements characterized UFKWs at two relatively narrow altitude bands, with one measurement altitude near 80 km and a second near 150 km.

Thermospheric density profiles from solar occultation (SO) measurements made by the Extreme Ultraviolet Monitor (EUVM) onboard MAVEN [2] provide a means to characterize multi-sol period waves, including UFKWs, as a function of altitude. In this study, we present first-ever observations of UFKWs as a function of altitude, including observations of rapid wave energy dissipation near 170 km indicative of wave breaking.

**Data:** MAVEN EUVM SO measurements are made using EUVM's 17-22 nm band, which measures $CO_2$ density from ~120-200 km [3]. Density profiles are found by first deriving the column density as a function of altitude from the measured extinction profile observed as the Sun sets or rises over the horizon as viewed from the spacecraft. Number density profiles are then found by applying an Abel Transform to the column density profiles. Measurements are made at either the dawn or dusk terminators (or both) if EUVM is sun-pointed as the spacecraft enters/leaves eclipse at a cadence of ~0.2 sol determined by the ~4.5 hour MAVEN orbital period.

We analyze approximately 60 sols of data measured at the dusk terminator beginning on 1 December 2014. For this time period, the observed latitudes were over the southern hemisphere and began near -67° and ended near the equator. Also, Mars was near perihelion and southern summer, with $L_s$ near 265° for this time period.

**Results:** Figure 1 shows time series of wave amplitudes at 5 altitudes. We find strong 10-15 K oscillations with periods between 2 and 4 days at all altitudes. We perform a wavelet analysis on time series of the data binned in 1 km bins and find a cessation of wave activity from sol 25 to 35 or, equivalently, between 35° and 18° South latitude. At 135 km, the dominant wave period is 3 sols for the first half of the dataset and 4 sols for the second half. The dominant wave periods are approximately 1 sol shorter for altitudes above ~150 km than they are at 135 km.

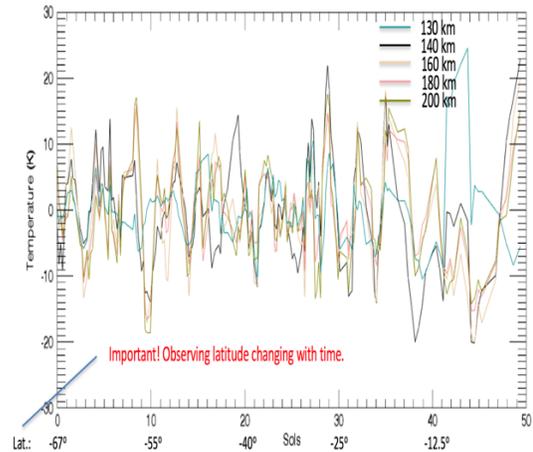

**Figure 1**. Time series of wave amplitudes at 10 km intervals from 130 km to 200 km.

It is also interesting to note that the wave amplitudes grow by ~5K in the latter half of the time range considered. This may be because these observations correspond with lower latitudes, allowing the waves to propagate more freely upward. However, given the strength of the wave amplitudes earlier on, when the observations were of the high latitude thermosphere, the theoretical prediction of UFKWs being equatorially trapped holds very weakly for this time period.

The periods of high wave activity correspond with periods of significant cooling in the lower thermosphere reported by [3] and attributed to thermostatic cooling triggered by strong 27-day solar variability occurring at the time. The enhanced wave activity may be a corollary of this thermostatic cooling. On the other hand, the apparent correlation between 27-day solar variability and cooler thermospheric temperatures may be the result of the latitudinal distribution of wave activity being aliased into the wave amplitude time series. We discuss both possibilities in our presentation.

If we assume the 2-3 day temperature oscillations are the result of gravity wave modes, which is valid for UFKWs, then the wave potential energy as a function of altitude and time can be computed from the EUVM SO density and temperature altitude profiles. The results of these calculations are shown in Figure 2, where the units are energy density ($J/m^3$). As expected, the cessation in wave activity between sols 25 and 35 corresponds with an abrupt drop in wave potential energy at all altitudes. Also evident from this figure is relative constant energy with altitude from 130 to 170 km, and an abrupt drop of wave potential energy near 170 km. This can be interpreted as being the result of waves freely propagating up to 170 km above which they rapidly dissipate their energy.

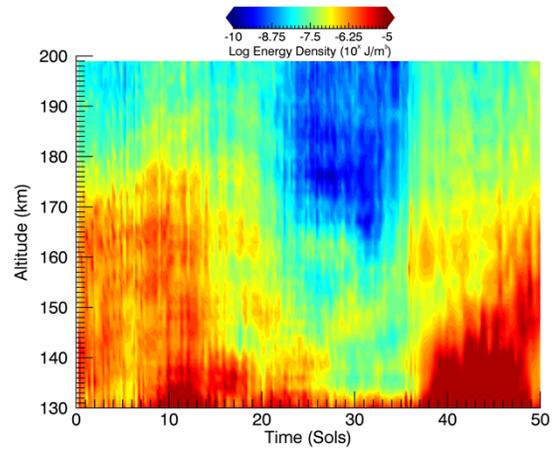

**Figure 2**. Potential energy of 2-3 day wave amplitudes.

**Conclusions:** MAVEN EUVM SO measurements provide a means to characterize long-period waves in the Mars thermosphere as a function of altitude, time and latitude. These observations show UFKWs as being the dominant source of day-to-day variability from low to high latitudes during Southern Summer. UFKWs propagate freely below 170 km, above which they rapidly dissipate.

**References:** [1] Gasperini, F. et al. (2018) *GRL, 45*. [2] Eparvier, F. et al. (2015) *Space Sci. Reviews, 195*. [3] Thiemann, E. et al. (2018) *JGR: Space 1234*.